\documentclass[referee]{aa}
\usepackage{graphicx}
\usepackage{txfonts}
\begin{document}

\title{Interstellar Extinction by Spheroidal Dust Grains}

 \author{Ranjan Gupta
            \inst{1}
    \and
    Tadashi Mukai
    \inst{2}
    \and
    D.B. Vaidya
    \inst{3}
    \and
    Asoke K. Sen\inst{4}
            \and
 Yasuhiko Okada\inst{2}
    }

     \offprints{Ranjan Gupta, \email{rag@iucaa.ernet.in}}

     \institute{IUCAA, Post Bag 4, Ganeshkhind, Pune-411007, India
               \and The Graduate School of Science and Technology,
       Dept. of Earth \& Planetary Sciences, Faculty of Science
       Kobe University, Nada, Kobe 657-8501, Japan
       \and Gujarat College, Ahmedabad-380006
               \and  Department of Physics, Assam University
       Silchar-788011, India}

     \date{}

     \abstract{

Observations of interstellar extinction and polarization indicate
that the interstellar medium consists of aligned non-spherical 
dust grains which show variation in the interstellar extinction curve
for wavelengths ranging from NIR to UV. To model the extinction
and polarization, one cannot use the conventional Mie theory which
assumes the grains as solid spheres. We have used a T-matrix based
method for computing the extinction efficiencies of spheroidal 
silicate and graphite grains of different shapes (axial ratios) and
sizes and used these efficiencies to evaluate the interstellar 
extinction curve in the wavelength range $3.4 - 0.1 \mu m$.
A best fit linear combination of silicate and graphite grains of
not very large axial ratio, fits the observed extinction curve 
reasonably well.

We calculate the volume extinction factor $\rm V_{c}$, which is
an important parameter from the point of view of the cosmic abundance,
for the spheroidal grain models that reproduce the interstellar
extinction curve. We find that the shape of the grains do not affect
the volume extinction factor.

Finally we have also studied the extinction and linear polarization
efficiencies for aligned spheroids. The results show that the shape
of grains affects the linear polarization efficiencies considerably
for various orientation angles of the spheroids. 

     }

     \maketitle
  %
  %________________________________________________________________

\section{Introduction}

  The exact solution to Maxwell's equations to calculate absorption,
  scattering and extinction of electromagnetic waves by homogeneous
  isotropic spheres is provided by Mie theory (Van de Hulst, 1981).
  Spherical grain models have been used by many authors to evaluate the
  interstellar extinction curve (e.g. Mathis et. al. 1977, Draine and Lee
 1984).
  However, the spherical grains can not explain the observed interstellar
  polarization that accompanies interstellar extinction.
   Mie theory would also not
  be strictly applicable if the particles have anisotropic optical
 properties,
  as for example graphite.  Using the method of separation
  of variables Asano and Yamamoto (1975) have given the analytical
  solutions for homogeneous isotropic spheroidal particles. The method
  of separation of variables is based on expanding the incident, internal
  and scattered fields in vector spheroidal functions. For spheroids with
  large sizes, large refractive indices
  or absorbing particles, the system of linear equations becomes large
  and it becomes a difficult mathematical and numerical problem.
  Recently, Voshchinnikov and Farafonov (1993) have improved
   the separation of variables method which is applicable to large
  aspect ratios. Several numerical techniques are now being in practice
  to obtain scattering and absorption cross sections for nonspherical
  particles (e.g. Bohren and Huffman 1983; Mishchenko et. al., 2000). 
The T-Matrix
  technique is based on expanding the incident field in vector spherical
  wave functions (VSWFs) regular at the origin and expanding the
  scattered field outside a circumscribing sphere of a scatterer in
  VSWFs regular at far field region. The T-Matrix transforms the expansion
  coefficients of the incident field into those of the scattered field
  and can be used to compute any scattering characteristic of a
  nonspherical particle (for details on T-Matrix method see Mishchenko
  et. al., 2002).

 One can also use the discrete dipole approximation (DDA see viz. Draine
 1988) to
 study the extinction properties of the nonspherical inhomogeneous
 particles (see Borghese et. al. 2003). We have used the DDA
 to study the extinction properties for porous and composite particles
 (Vaidya et. al. 1997, 1999, 2001;  Paper I, II and III hereafter).
 However, the DDA method required considerable computer time and memory.
 The T-Matrix code on the other hand (Mishchenko et. al., 2002)
 runs much faster and the results obtained can be tuned
 with much ease since the input parameters to the code can be adjusted
 and re-run in a short time. However, the T-Matrix code in its present
 form cannot be used for studying inhomogeneous (e.g. porous, fluffy,
 composite) particles (Mishchenko et. al., 2002).

 In the section 2 we present the spheroidal grain models and describe
 the T-Matrix method; in section 3 we give the results of our
 calculations; i.e. extinction curves as functions of grain size, shape
 and wavelengths, the model interstellar extinction curves, and comparison
 of these model curves with the observed interstellar extinction curve
 (Savage and Mathis 1979, Whittet 2003) and the influence of particle
 shape on the determination of interstellar extinction and
 in the section 4 we provide the conclusions of the present study.

 \section{T matrix method and Spheroidal grain models}

  The T-Matrix method was first introduced by Waterman (1965)
  for studying electromagnetic scattering by single, homogeneous
  nonspherical particles. The standard scheme for computing the
  T-Matrix for single homogeneous scatterers in the particle reference
  frame is called the extended boundary condition method (EBCM)
  and is based on the vector Huygens principle (Waterman 1971).
  The general problem is to find the field scattered by an object bounded
  by a closed surface. The Huygens principle establishes the relationship
  between the incident field, the total external field (i.e. the sum of
  the incident and the scattered fields), and the surface field.
  Technically the incident and the scattered waves are expanded
  in regular and outgoing vector spherical wave functions (VSWFs);
  the concept of  expanding the incident and the scattered waves in
  VSWFs and relating these
  expansions by means of  T-Matrix has proved very useful.
  The great computational efficiency of the T-Matrix approach
  to study electromagnetic
  scattering by nonspherical particles with various shapes and sizes
  has found numerous applications. Several FORTRAN T-Matrix
  codes are available on the world wide web at
  http://www.giss.nasa.gov/$\sim$crmim. The codes
  compute the complete set of scattering characteristics, the optical
 cross sections, expansion coefficients, and the scattering matrix
  for randomly oriented and aligned particles.

 In the present study the axial ratios 
(ratio of horizontal to rotational axes hereafter called
as AR) of the spheroidal grains ranging from 0.5 to 2.0 have been
 taken. Please note that for prolates we have AR $<$ 1 and for
 oblates we have AR $>$ 1. The aspect ratio $\epsilon$ is the
 ratio of the larger and smaller axes as defined by Mishchenko et. al..
 (2002) and thus both a prolate with AR=0.5 and an oblate with AR=2.0
 will have the aspect ratio $\epsilon$=2. For spheres AR=1.0
 and $\epsilon$=1.

The observed interstellar polarization requires that
interstellar grains be aligned and nonspherical in shape. 
Hence, in addition to the calculations on
random orientations (with orientation averaging) of the spheroidal
grains we have also performed calculations for the aligned spheroids
at several orientation angles.

  \section{Results:}

 {\bf 3.1 Extinction properties of Spheroidal Grains}

  One of the main objectives of the paper is to study the extinction
 efficiencies of the spheroidal silicate and graphite grains, obtained
 using the T-Matrix calculations and see the effects of shape of the
 grains i.e. spheres, oblates and prolates and various sizes
 which are typical to interstellar grain sizes i.e. 0.001 to
 about 0.500$\mu$.
 These materials (i.e. silicate and graphite) have been the ingredients
 for most of the spherical grain models (see for example
 Mathis et. al. 1977, Draine and Lee 1984, Weingartner and Draine 2001).
 We present the results for the spheroidal grain models;
 i.e. the extinction efficiencies $\rm Q_{ext}$ as a function
 of wavelength in the spectral range (3.4 to 0.1 $\mu m$).

Silicates:

 The Figure 1 shows the extinction efficiency $\rm Q_{ext}$
 for silicate grains for four grain
 sizes i.e. 0.01, 0.05, 0.1 and 0.2 $\mu$.
 The y-axis in these plots are in the form of an extinction efficiency
 ratio $\rm Q_{ext}(Spheroid)/Q_{ext}(Sphere)$ to bring out
 the effect of oblates and prolates as compared to spheres.

 \begin{figure}
 \includegraphics[width=\textwidth]{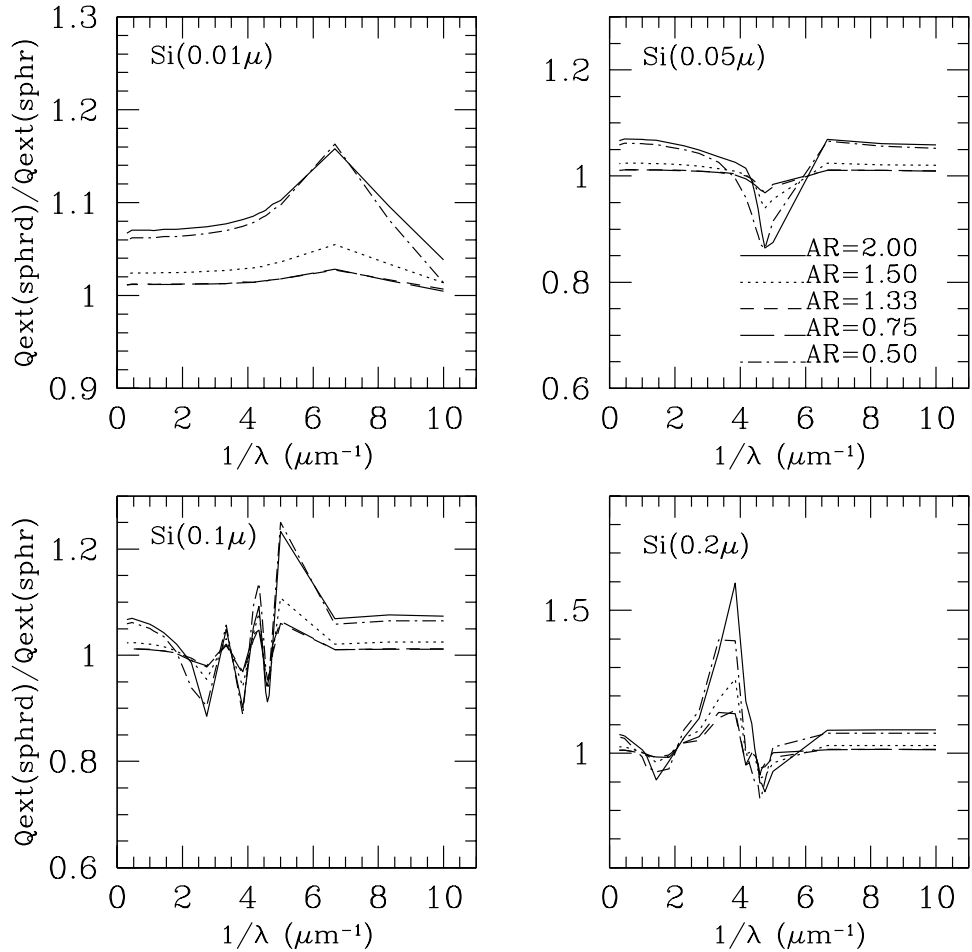}
 \caption{Extinction efficiencies
 ($\rm Q_{ext}(Spheroid)/Q_{ext}(Sphere)$)  versus wavelength
 for randomly oriented
spheroidal silicate grains for four different grain sizes
 and five different axial ratios.}
 \end{figure}

 \begin{figure}
 \includegraphics[width=\textwidth]{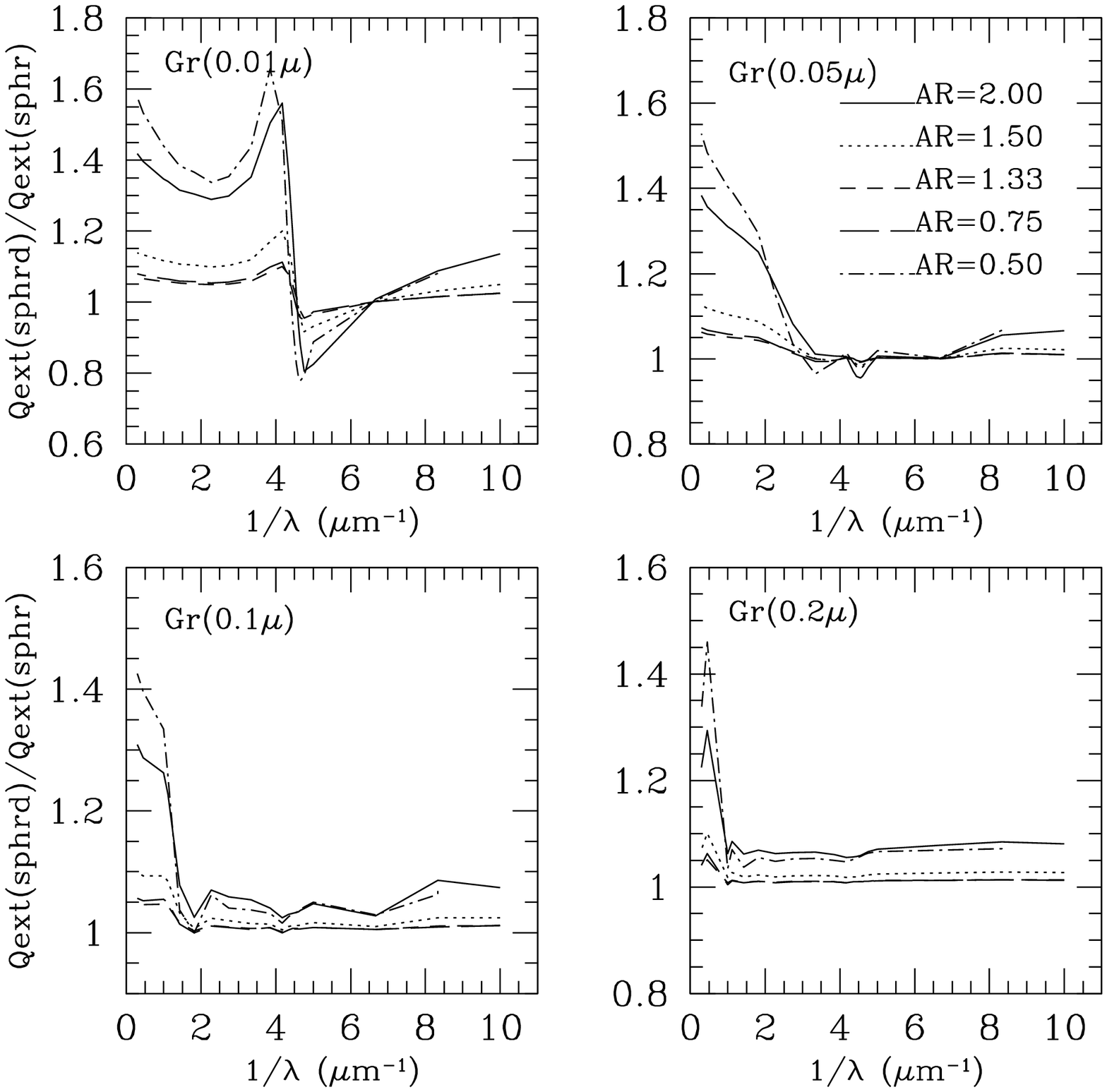}
 \caption{Extinction efficiencies
 ($\rm Q_{ext}(Spheroid)/Q_{ext}(Sphere)$)  versus wavelength
 for randomly oriented spheroidal graphite grains for 
four different grain sizes and five different axial ratios.}
 \end{figure}

It is seen from this figure that the grains with the same
aspect ratio $\epsilon$=2 but having different axial ratios,
AR=0.5 (Prolates) and AR=2.0 (Oblates), display similar
extinction properties in the entire wavelength range,
i.e 3.4 - 0.1 $\mu m$. Results for the grains with other aspect ratio
$\epsilon$ =1.33, having an axial ratio AR=0.75 (Prolates) and
AR=1.33 (Oblates) also show similar extinction property.
These results on spheroids are consistent with the results
reported by Mishchenko et. al. (2002).
It is also seen that the extinction efficiency ratio
$\rm Q_{ext}(Spheroid)/Q_{ext}(Sphere)$
for the silicate grains with aspect ratio $\epsilon$=1.33 (i.e.
AR=0.75 and 1.33) is close to AR=1.0 (sphere),
i.e. there is no appreciable variation in the extinction efficiency from
spheres;
whereas spheroidal grains with aspect ratio $\epsilon$=2 
(i.e. AR=0.5 and 2.0),
show considerable variation in the extinction from sphere i.e. the ratio
$\rm Q_{ext}(Spheroid)/Q_{ext}(Sphere)$ deviates from 1.0.

Graphites:

Figures 2 shows extinction curves for spheroidal graphite grains.
For small graphite grains (a=0.01 $\mu$) the '2175\AA~' extinction
feature is displayed by all the spheroidal grain models, but the
peak is shifted for various axial ratios.
In order to emphasize this aspect we show in Figure 3
the extinction efficiency $\rm Q_{ext}$ for the spheroidal 
graphite grains for a=0.01 and 0.05 $\mu$ in the wavelength range 
0.25 - 0.20 $\mu m$ (i.e. 4 - 5 $\mu m^{-1}$).
These curves clearly show the shift in the extinction peak with
the shape of the grains (i.e. axial ratio). For a=0.01 $\mu$ the 
extinction peak at 4.6$\mu m^{-1}$ for spherical
grains (AR=1 and Mie) shifts to 4.4$\mu m^{-1}$ for the spheroidal
grains with AR=2.0. These curves also show the variation in the
width of the feature. These results on the spheroidal grains indicate 
that the shape of the grains plays an important role in studying 
extinction properties and needs to be studied in more details.
Voshchinnikov (1990) has also found the variation in the 
'2175\AA~' feature with the shape of the grain.
In our earlier study on the porous grains 
(Vaidya et. al. 1997 \& 1999 i.e. Paper I and II)
we had found the shift in the the '2175\AA~' peak as well as 
the variation in the width of the feature for the graphite
grains with porosity. Draine and Malhotra (1993) have found the shift
in the central wavelength of the feature in the coagulated graphite 
grains but did not find any appreciable change in the width.
It is to be noted  from Figure 2 that for spheroidal graphite
grains with aspect ratio $\epsilon$=1.33 (i.e. AR=0.75 and 1.33), there is
no appreciable variation in the extinction from spheres.
It is also seen that for spheroidal graphite grains
with a$>0.05\mu$, the extinction does not deviate much
from spheres in the optical and the UV spectral range.
Please note that for prolate graphite with AR=0.5 in Figure 2
for a=0.01 and 0.05$\mu$, the T-Matrix calculation does not converge
beyond 8 $\mu m^{-1}$.

 \begin{figure}
 \includegraphics[width=\textwidth]{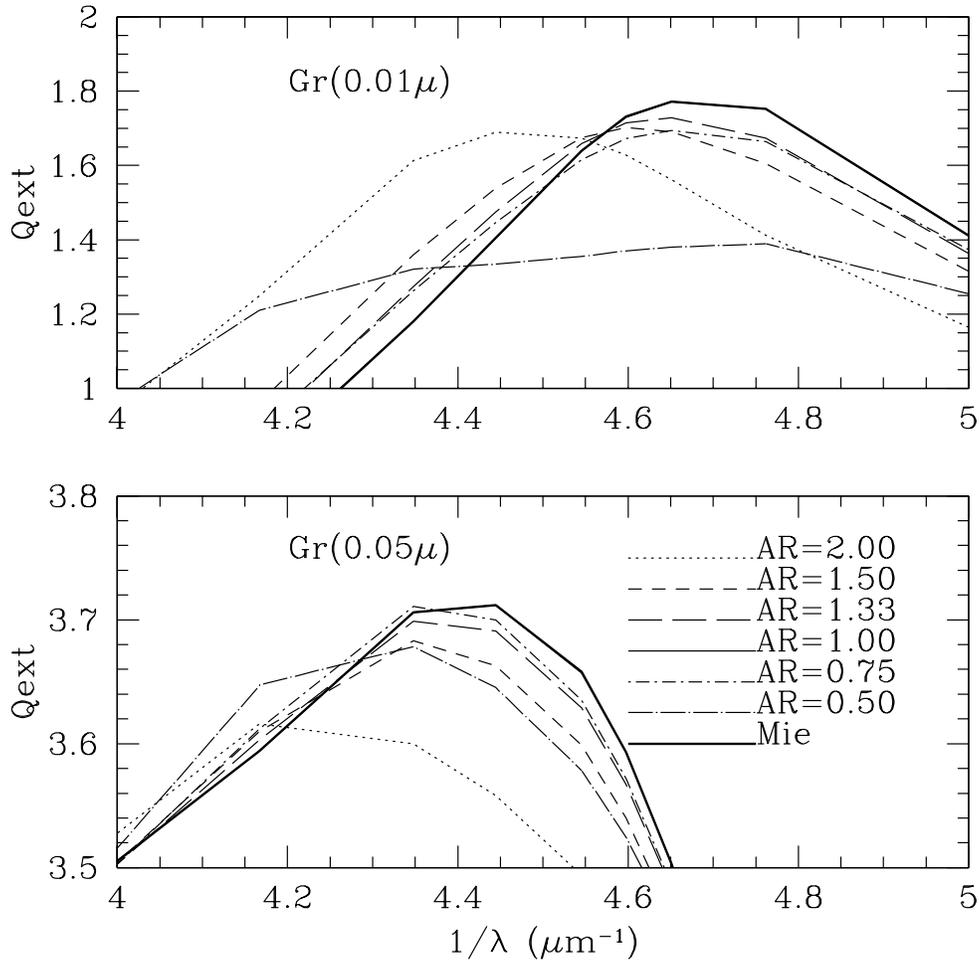}
 \caption{Extinction efficiencies
 $\rm Q_{ext}$ versus $1/\lambda$ for randomly oriented
graphite grains of sizes
a=0.01 and 0.05 $\mu$ in the UV peak region for various
axial ratios. The axial ration AR=1.00 and Mie curves overlap
each other as expected. The shift in the peak extinction is clearly
seen for various axial ratios.}
 \end{figure}

\newpage

{\it Extinction and Polarization efficiency for Aligned Spheroids}

As mentioned earlier the existence of interstellar
polarization requires that the grains
must be nonspherical and aligned (see e.g Greenberg 1968,
Wolff et. al. 1993, Voshchinnikov \& Farafonov 1993).
We have calculated the extinction and linear polarization for
the aligned spheroidal silicate
and graphite grains at several orientation angles.
Figures 4(a-d) show $\rm Q_{ext} = [Q_{ext}(E) + Q_{ext}(H)]/2$ and 
linear polarization efficiency 
$\rm |Q_{pol}| = Q_{ext}(E) - Q_{ext}(H)$ 
for aligned (at a fixed orientation angle $\beta = 45^{\circ}$)
oblate spheroids for a = 0.1 $\mu$ in the wavelength region
$\rm 3.4 - 0.3 \mu m$; where $\rm Q_{ext}(E)$ and $\rm Q_{ext}(H)$
are extinction efficiency factors for the directions of the incident
field vector Q(E) and perpendicular Q(H) to the axis of the spheroid.
These results on the aligned oblates show that there is no appreciable
variation in the extinction but the polarization shows considerable
variation with the axial ratio. 
Figs 4(e-h) show the extinction and linear polarization efficiency
for a fixed axial ratio AR=1.33 but with varying orientation 
angles $\beta$ from 0 to 90$^{\circ}$ for oblates. Here again the 
extinction does not show any appreciable change but the polarization 
does show a lot of variation.
The aligned oblate silicate spheroids with AR =1.33 produce maximum
linear polarization between 2 and 3 $\mu m^{-1}$, whereas the aligned
graphite grains produce the maximum polarization between 
1 and 2 $\mu m^{-1}$. These results on linear polarization are 
consistent with the results obtained for spheroids by 
Wolff et. al. (1993).
Figs 5(a-h) show extinction and linear polarization efficiencies for
the aligned prolate silicate and graphite spheroidal grains.
The results in 5(a) and (b) on aligned prolates do not show 
appreciable variation with the axial ratio. However, the extinction 
varies with the orientation angles in 5(e) and (f) for aligned prolates.
Note that the polarization efficiency of oblate silicates in Figures
4 (c \& g) and prolate silicates in Figures 5 (c \& g) show considerable
differences but rest of the curves for oblates and prolates follow
similar trends.

\begin{figure}
\includegraphics[width=\textwidth]{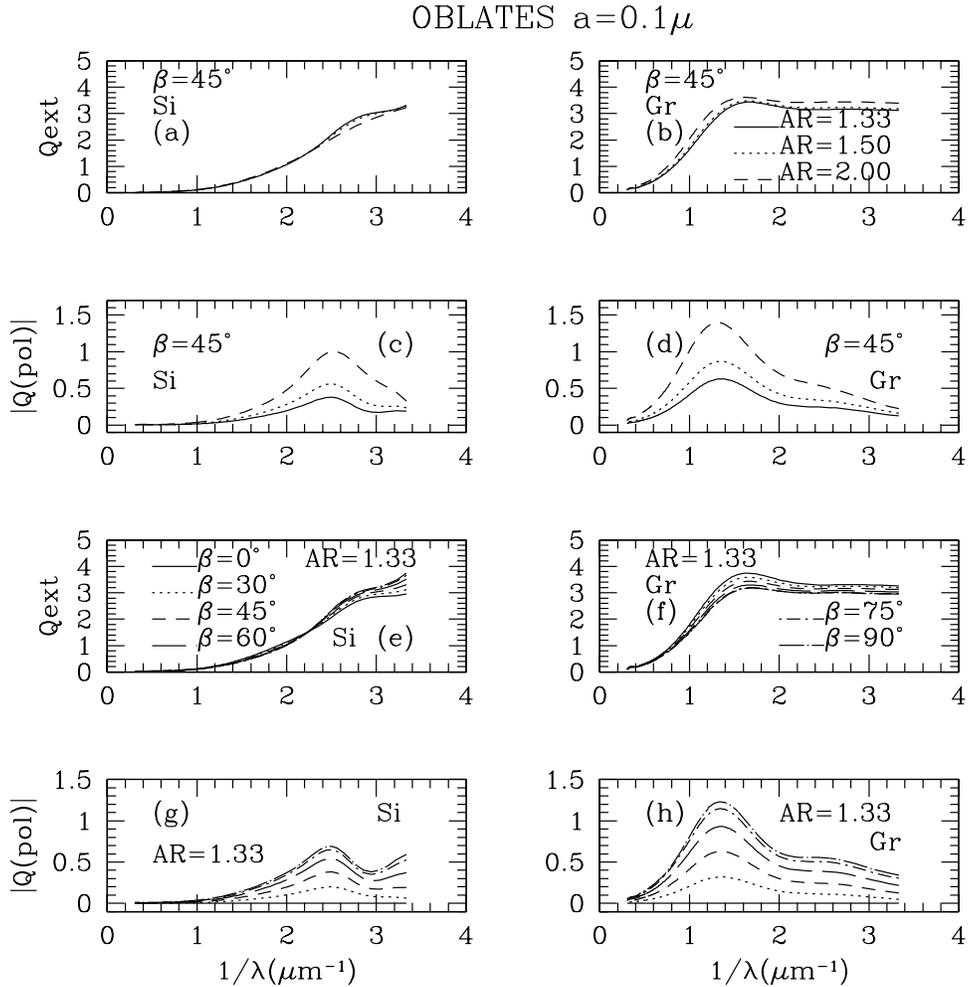}
\caption{Extinction $\rm Q_{ext}$ and  linear polarization 
efficiencies versus $1/\lambda$
for aligned oblate spheroids (Silicates and Graphites with
grain size a=0.1$\mu$). 
The panels (a), (b), (c) and (d) are
for a fixed orientation ($\beta$=45$^{\circ}$) and various
axial ratios AR of the aligned spheroids. 
The panels (e), (f), (g) and (h) are for a fixed axial ratio
AR=1.33 but with various orientation angles $\beta$.}
 \end{figure}

\begin{figure}
\includegraphics[width=\textwidth]{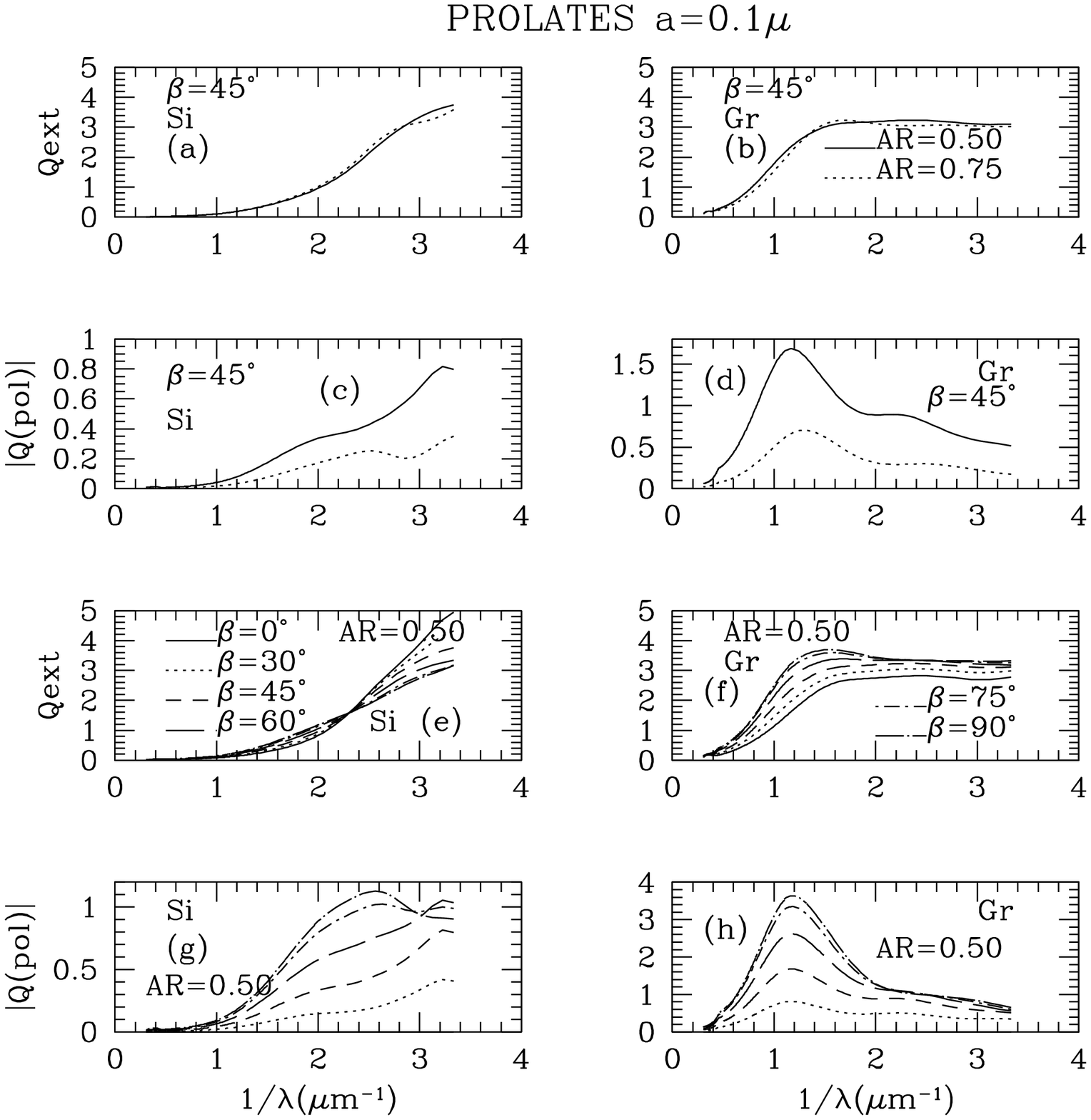}
\caption{Extinction $\rm Q_{ext}$ and  linear polarization efficiencies
versus $1/\lambda$
for aligned prolate spheroids (Silicates and Graphites with
grain size a=0.1$\mu$). 
The panels (a), (b), (c) and (d) are
for a fixed orientation ($\beta$=45$^{\circ}$) and various 
axial ratios AR of the aligned spheroids. 
The panels (e), (f), (g) and (h) are for a fixed axial ratio
AR=1.33 but with various orientation angles $\beta$.}
 \end{figure}

 {\bf 3.2 Interstellar Extinction Curve}

  We use the extinction efficiencies of the silicate and graphite grains
  and the power law
  grain size distribution (i.e. $\rm n(a) \sim a^{-3.5}$, MRN size
 distribution (see Mathis, et. al., 1977 and Paper I, II \& III)
  to reproduce the average observed interstellar extinction curve
 (Savage and Mathis 1979, Whittet, 2003). We have evaluated the
 interstellar extinction curves for grain size distribution with
the smaller size limit starting at a=0.001$\mu$ and the larger size limit
 up to a=1.0$\mu$. Normalized interstellar extinction curve
 is generated for both silicates and graphites and then combined
 linearly as described in our Papers I, II and III, to obtain the
 best fit with the observed extinction curve using a $\chi^2$ minimizing
scheme.

 Apart from the various sizes tried for the lower and upper
 limits of the size distribution; the power law index of -3.5 was
 also changed from -2.5 to -4.5 in steps of 0.5 and the value of -3.5
 seems to be the best of the lot for obtaining best fits to the
 observations. Finally calculations were also repeated
 with different axial ratios and the best fit value of 
$\chi^2=0.012837$ was obtained with AR=1.00 i.e. spheres as was 
predicted by the MRN model. However, the spherical grains are not 
realistic, and besides, the observed interstellar polarization
requires the interstellar grains to be non-spherical.
Thus we keep the oblate spheroids with AR=1.33, for which
the best fit $\chi^2$ value is 0.013210, as the model for interstellar
extinction.

The Figure 6 shows the best fit $\chi^2$ minimized model curve along with
 the observed interstellar extinction curve.
 The top panel show the full curve in the wavelength region
 3.4 to 0.1 $\mu m$ i.e. far UV to NIR. The bottom panel highlights
 the UV bump region. The best model curve thus consists of a size
 distribution a=0.005-0.250 $\mu$ in steps of 0.005 $\mu$
 and with spheroids with AR=1.33.
 The observed points (shown as filled square dots)
 have been evaluated at wavelength steps of 50\AA~ by interpolation
 since the original data (Savage and Mathis 1979, Whittet, 2003)
 has much fewer wavelength points though it covers this large range.
 Similarly the T-Matrix model calculations were specially
 performed at this high wavelength resolution to provide smoother
 interstellar curves. Such calculations are feasible with T-Matrix due to
 its intrinsic property for fast computations.

  \begin{figure}
 \includegraphics[width=\textwidth]{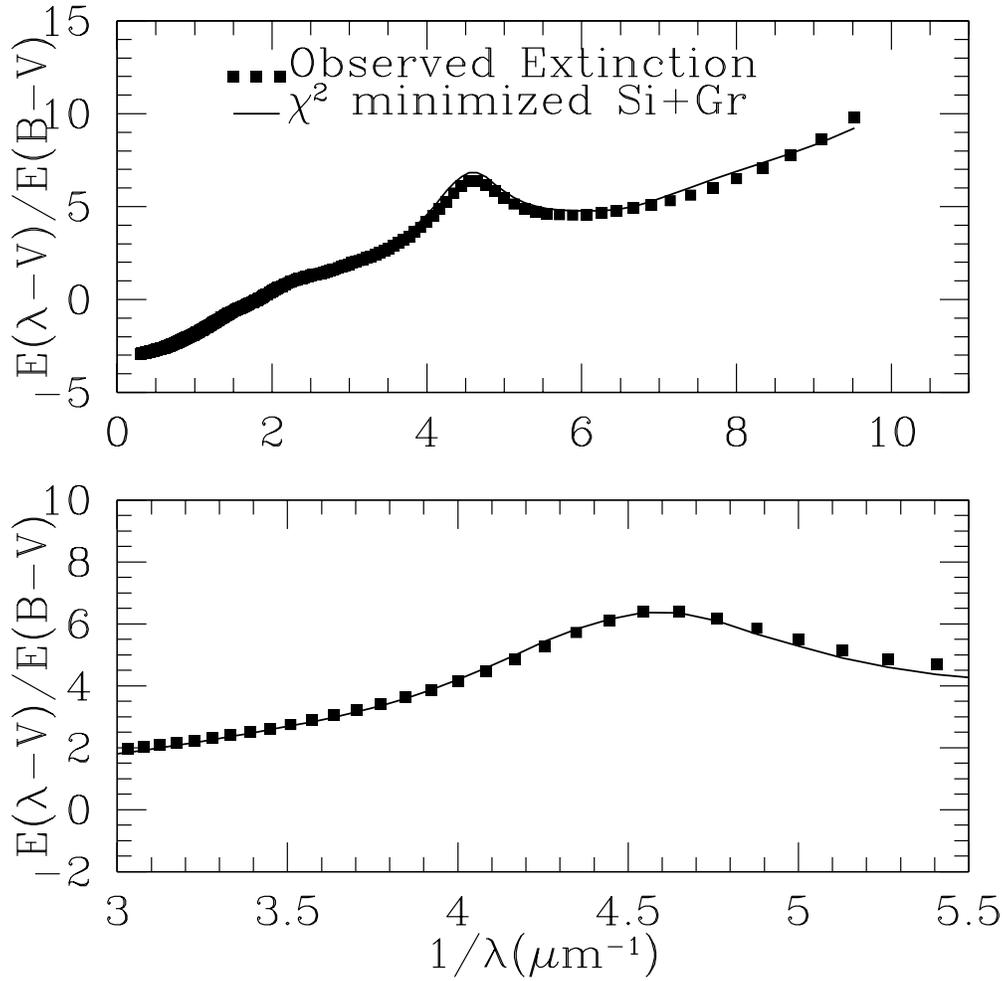}
    \caption{The top panel shows the comparison of observed interstellar
 extinction curve with the best fit model combination curve for
 spheroidal silicate and graphite grains (AR=1.33)
in the wavelength region of
 3.4 to 0.1 $\mu m$ using a grain size distribution of 0.005-0.250 $\mu$
 in steps of 0.005 $\mu$.
 This plot has a wavelength resolution of 50\AA~
for the model values and also for the observed values though the observed
 values are available at much lower resolution and at much fewer
 wavelength points. The lower panel highlights the UV bump region
 of the interstellar curve fitting.}
  \end{figure}

{\bf 3.3 Volume Extinction factor and Abundance constraints}

An important parameter from the point of view of cosmic abundance
is the volume extinction factor $\rm V_{c}$ which is
defined as the ratio of total volume of the grains to the total 
extinction at a given wavelength i.e. $\sum V/\sum C_{ext}(\lambda)$
(see for example, Greenberg 1968, Vaidya et. al. 1984). 
$\rm V_{c}$ directly determines the amount of material required to 
render the level of extinction at a specific wavelength.
Table 1 shows the volume extinction factors $\rm V_{c}$ at $\lambda
=0.55 \mu m$ for the spheroidal grain
models that reproduce the observed interstellar extinction (Figure 6).

\begin{table*}
\begin{center}
\caption{Volume Extinction factor $\rm V_{c}$ for spheroidal grains.}
\begin{tabular}{lccc} \hline \hline
Shape & Axial ratio & Silicate & Graphite\\
\hline
Oblate & 1.33 & 0.266 & 0.102\\
       & 1.50 & 0.267 & 0.100\\
       & 2.00 & 0.271 & 0.092\\
\hline
Sphere & 1.00 & 0.265 & 0.111\\
\hline
Prolates & 0.75 & 0.266 & 0.103\\
         & 0.50 & 0.272 & 0.093\\
\hline
\end{tabular}
\end{center}
\end{table*}

Results in Table 1 indicate that the variation in $\rm V_{c}$ due to 
shape (oblates or prolates) for both silicate and graphite grains 
is less than 10 \% , hence the atomic-abundance constraints based on 
values of $\rm V_{c}$ for either spheres or spheroids of each material 
would be identical. These results showing effect of shapes on volume
extinction factor are consistent with the results obtained by 
Greenberg (1978).

\section{Summary and Conclusions}

The extinction efficiencies for prolates, oblates
 and spheres are obtained using the T-Matrix calculations for silicate 
and graphite grains (of different sizes) in the wavelength range of 
3.4 - 0.1 $\mu m$.
 The study brings out the effect of the shape of the grains
 on the extinction and linear polarization.  Our results show that the
shape of the grain plays an important role in determining the extinction
properties of the grains and requires a more detailed study with other
shapes and axial ratios. The
 spheroids with same aspect ratios ($\epsilon$) tend to show similar
 deviations in $\rm Q_{ext}$ values when compared with the $\rm Q_{ext}$
 values of spheres and this phenomena has been noted by
 Mishchenko et. al. (2002).

 The $\rm Q_{ext}$ computations were used to model the interstellar
 extinction and then compared with the observed curve. Best
 fits are obtained for grain size distribution a=0.005-0.250$\mu$ in
 steps of 0.005$\mu$ (with the power law exponent value of -3.5)
 with spheroidal grains having axial ratio of AR=1.33.

 The bare graphite/silicate grain
 model which we have used in the present study is consistent
 with observations of interstellar extinction.
The volume extinction factors
determined for the spheroidal grain models do not show appreciable
variation from spherical models, so the abundance constraints
for spheroidal grains and spheres would be identical.
  It should be noted the bare graphite/silicate
 grain model requires higher abundances
 of carbon and silicon than are recently available
 values for the interstellar medium (Mathis 2000).
 The real interstellar grains might be
more complex (composite, porous; see e.g. Mathis 1996, Vaidya et. al. 2001).

It is to be noted here that the T-Matrix method
to study the optical properties of the spheroidal silicate and
graphite grains is a new approach and there is enough scope for varying
the grain shape, size and composition for obtaining better model fits to
the observations. 
Recently composite grains have been used by Moreno et. al. (2003)
to model Comet Hale-Bopp dust grains. Petrova et. al. (2000) have
used modified version of T-Matrix to study grains and have
applied these results to model cometary grains.

 \begin{acknowledgements}
 The authors wish to thank Mike Mishchenko for the T-Matrix code.
 The comments from the referee Dr. N. Voshchinnikov were extremely useful
 to improve the quality of the paper. The authors also wish to thank
 DST, New Delhi, India and the Indo-Japan Co-operative Science
 Programme operated by JSPS, Tokyo, Japan for providing funds to RG
 and AKS for this collaboration.
\end{acknowledgements}

\end{document}